\documentstyle[12pt]{JHEP3}
\title{M5-branes with $3/8$ supersymmetry in  pp-wave background 
\thanks{e-mail: singh@physik.uni-halle.de}}
\author{Harvendra Singh\\
Fachbereich Physik, Martin-Luther-Universit\"at Halle-Wittenberg,
\\ Friedemann-Bach-Platz 6, D-06099 Halle, Germany}
\abstract{ 
We construct M5-branes with $3/8$ supersymmetry 
and find  that they preserve exactly half of the 
background pp-wave supersymmetries.
We explicitly write down the standard as well as supernumerary Killing spinors
and find that their respective
 numbers are also half of those for the pp-wave 
background. This is in line with the recent work of Dabholkar et.al. 
which shows that half-supersymmetric D-branes can be constructed in 
pp-wave backgrounds.}
\begin{document}
\def\be{\begin{equation}} \def\ee{\end{equation}}
\def\bea{\begin{eqnarray}} \def\eea{\end{eqnarray}} \def\ba{\begin{array}}
\def\ea{\end{array}} \def\ben{\begin{enumerate}} \def\een{\end{enumerate}}
\def\nab{\bigtriangledown} \def\tpi{\tilde\Phi} \def\nnu{\nonumber}
\def\lll{\label}
\newcommand{\eqn}[1]{(\ref{#1})}
\def\cC{{\cal C}} 
\def\cG{{\cal G}} 
\def\cd{{\cal D}} 
\def\a{\alpha}
\def\b{\beta}
\def\g{\gamma}\def\G{\Gamma}
\def\d{\delta}\def\D{\Delta}
\def\ep{\epsilon}
\def\e{\eta}
\def\z{\zeta}
\def\t{\theta}\def\T{\Theta}
\def\l{\lambda}\def\L{\Lambda}
\def\m{\mu}
\def\f{\phi}\def\F{\Phi}
\def\n{\nu}
\def\p{\psi}\def\P{\Psi}
\def\r{\rho}
\def\s{\sigma}\def\S{\Sigma}
\def\ta{\tau}
\def\x{\chi}
\def\o{\omega}\def\O{\Omega}
\def\k{\kappa}
\def\pa {\partial}
\def\ov{\over}
\def\br{\nonumber\\}
\def\ud{\underline}
\section{Introduction}
Hpp-waves are 
maximally supersymmetric plane-fronted parallel wave configurations of type 
IIB string theory \cite{blau,blau1} which can be obtained by taking the 
Penrose 
limit \cite{penrose,gueven} of the maximally supersymmetric $AdS_5\times 
S^5$ spacetime.  
Note that unlike Minkowski spacetime, pp-waves are asymptotically non-flat geometries, 
however, string theory in these
backgrounds is exactly solvable \cite{matsaev,matsaev1}
 and have important consequences for
dual conformal field theories \cite{maldacena}. 
Several quick advances have taken place in the following works 
\cite{bmn,blau2,mukhi,alish,gomis,zayas,billo,kim,takaya,nunez,
das,chu,clp,beren,dabhol,
hull,lee,lu1,hatsuda,kumar}. 
There  also have
been recent works exposing  various possible pp-wave solutions in 
string theory \cite{clp,hull}.  

Usually embedding of extended 
branes in a Minkowskian background breaks half of the background supersymmetries. 
For  D-brane in pp-wave background to have $1/2$
supersymmetry brane must be embedded in maximally supersymmetric Hpp-wave 
backgrounds. Recently in a paper Dabholkar et. el. \cite{dabhol} 
have proposed existence of 
$1/2$ supersymmetric Dp-branes (for $p=3,5$ and 7) in Hpp-wave 
backgrounds, for subsequent work see \cite{skend}. 
The existence of D-branes in pp-wave backgrounds definitely leads 
to the existence of  M2 and M5-branes in eleven dimensional M-theory. 
Our interest in this paper is to find out 
supersymmetric 11-dimensional branes in pp-wave backgrounds. 
In particular we  are looking for 
$3/8$ supersymmetric M5-brane embedded in a $3/4$ supersymmetric 
11-dimensional pp-wave background. The reason for existence for such 
branes is 
provided by the existence of M5-branes in 
$AdS_3\times S^3\times T^5$ background. Note that $AdS_3\times S^3\times T^5$ 
preserves only $1/2$ of the supersymmetries  while  the Penrose 
limit  of  it leads to the pp-waves with 
$3/4$ supersymmetry \cite{clp}. 
Thus, in general, there is enhancement of the supersymmetries in the 
Penrose limit. 

The paper is organised as follows. The section-2 is a quick review of 
the 
general properties of pp-wave solutions with some specific  
details which will be needed for the  section-3. In the section-3
we obtain  supersymmetric M5-brane in pp-wave background. These 
five-branes are circularly symmetric or have smearing along one of the 
transverse coordinates.  We also construct explicitely the  
the Killing spinors. In particular we find that the `standard' as well as the 
`supernumerary' Killing spinors are exactly halved by an embedding of the 
five-brane in the pp-wave background. There is no mixing between the two kind 
of spinors. We also discuss the spacetime dependences of the Killing spinors.    

\section{Review: pp-waves and the traceless matter}

As explained in \cite{blau1} the Penrose limits \cite{penrose,gueven} of 
anti-de Sitter spacetimes $
{AdS_m\times S^n\times}$ $ S^p\times T^q$ along a null geodesic with 
generic orbit (i.e. with a non-zero component along a sphere) leads to
a pp-wave metric (for $m,n\ge2$) 

\bea\label{1a}
ds^2= dudv + \rho^2 sin^2({u\over 2\r}) \sum_{\m=1}^{m-1}(dy^\m)^2+ 
sin^2({u\ov2}) 
\sum_{\a=1}^{n-1}(dy^\a)^2 + \sum_{k=1}^{p+q}(dz^k)^2
\eea
where $\r$ is a parameter.
The pp-wave solution \eqn{1a} is written in Rosen coordinates and depends only on 
light-cone coordinate 
$u$. However, one could face a more generic situation of the following 
type
\bea
ds^2= dudv +\sum_{a=1}^{m+n-2} ({sin(\r_au)\over2\r_a})^2 (dy^a)^2+ 
\sum_{k=1}^{(p+q)}(dz^k)^2 .
\eea
{}From here we can switch to the new set of coordinates $(dx^+, 
dx^{-},x^a,z^k)$  
\bea
x^{-}=u/2,~~~x^+=v-{c\ov4}u -{1\ov4} \sum_a {sin(2\r_a u)\over2\r_a} y^a 
y^a, ~~~
x^a= y^a {sin(\r_au)\over2\r_a},~~~z^k=z^k ,\br
\lll{3a}\eea
which are slightly different from those given in \cite{blau1}
as we have included an arbitrary constant $c$.
In new coordinates we get the familiar form of the pp-wave metrics (or 
Cahen-Wallach spacetimes)
\bea\lll{wave}
&& ds^2= 2dx^+dx^{-} +W  (dx^{-})^2+ \sum_{a=1}^{m+n-2}(dy^a)^2+
\sum_{k=1}^{p+q}(dz^k)^2,\br
&&W=c-4\sum_{a=1}^{m+n-2}\r_a^2 (y^a)^2
\eea
where $c$ is an arbitrary positive constant which later on
 will be set equal to one.
 Note that for  metric \eqn{wave}  the nonvanishing 
componant of the Ricci 
tensor  is $R_{--}$ while the curvature scalar is vanishing.   
Therefore the 
Einstein's equation cannot 
be satisfied without  matter fields. The choice of 
the matter fields has to be 
such that the  energy momentum tensor is traceless ($T^M_{~M}=0$) as
curvature is vanishing. 
It has been shown in \cite{clp,hull} that  
special choices of the parameters $\r_a$ and the matter fields
give rise  to many 
supersymmetric pp-waves solutions in string theory.
Further the choice of 
$W$ in \eqn{wave} could still be of more general form.
For example 
$W$ could be taken as
\bea
 W=c+f_-(x^{-}) + [H_1(y^{\a_1})-m_1 
\sum_{\a_1=1}^{d_1}(y^{\a_1})^2]
+[H_2(y^{\a_2})-m_2 \sum_{\a_2=1}^{d_2}(y^{\a_2})^2]+\cdots
\eea
where $H_i$'s are the harmonic functions over the respective
Euclidean planes and $f_-(x^{-})$ is an arbitrary function of 
$x^{-}$. For such a wave solution the Ricci tensor becomes
\be
R_{--}=m_1 d_1+m_2 d_2 +\cdots \ ,
\ee
where $d_i$ are the dimensionalities of the respective homogeneous Euclidean 
coordinate patches. When parameters $m_i$ are chosen such that $R_{--}$ 
vanishes we obtain
pure gravitational pp-waves \cite{horowitz} which are Ricci flat and 
without matter fields. But in this paper we restrict ourselves to
$m_i\ge0$ and $R_{--}$  nonvanishing.  

\subsection{$3/4$ Supersymmetric pp-wave background} 
We consider the supersymmetric case of 11-dimensional pp-wave solution which 
follows 
by taking the appropriate Penrose limit of half-supersymmetric 
$AdS_3\times S^3\times T^5$ 
supergravity solution. Corresponding pp-wave is given by
\cite{clp} 
\bea\lll{wave1aa}
&& ds^2_{11}= 2dx^+dx^{-} +W~  (dx^{-})^2+ \sum_{a=1}^{4}(dx^a)^2+
\sum_{\a=5}^{8}(dy^\a)^2+(dy^9)^2.\br
&&G_4=2 m~ dx^{-} 
(dx^1dx^2dy^9+dx^3dx^4dy^9) ,~~~~W=c-m^2\sum_{a=1}^{4} (x^a)^2~ .
\eea
It has been shown in \cite{clp,hull} that above wave solution 
preserves 24 supersymmetries. Thus, in general,  pp-wave limits of the $AdS$
spacetimes are accompanied with the  enhancement of the supersymmetries. 

Let us now focus on the Killing spinors for the pp-wave background \eqn{wave1aa}.
These Killing spinors have been worked out in \cite{clp,hull}. We write 
them here 
more explicitely  as we shall require them in the next section.
We write down the tangent space metric as $ds^2=2e^{+}e^{-}+ 
e^a~e^a+e^\a~e^\a+e^9e^9$,
where tangent space indices are taken same as the space-time indices. 
The basis elements are given by 
$e^+=dx^{+}+(W/2) 
dx^{-},~e^{-}=dx^{-},~e^a=dx^a,~e^\a=dy^\a,~ e^9=dy^9$. It is easy to find 
that only non-vanishing spin connections are $\o^{+a}={1\ov2}\partial_a W 
dx^{-}$.
Then the Killing spinors are obtained by solving the following supersymmetry 
variations 
\be
\delta \Psi_M= \nabla_M\epsilon 
-{2\ov(4!)^2}(G_{PQRS}\G^{PQRS}_{~~~~~M}-8G_{MNPQ}\G^{NPQ})~\ep=0~.
\label{killing}\ee
For above background these reduce to the following set of equations
\bea
&& \pa_{+}\ep=0,~~~\br
&&[\pa_{-}+{1\ov4}\partial_a W 
~\g_{+}\g^a+{m\ov6}\g_9\Theta(\g_{+}\g_{-}+1)]\ep=0\br 
&&[\pa_a -{m\ov 12} \g_9(3\Theta\g_a-\g_a\Theta)\g_{+}]\ep=0\br 
&&[\pa_\a -{ m\ov 6} \Theta\g_\a\g_{+}]\ep=0\br 
&&[\pa_9 -{ m\ov 3} \Theta\g_9\g_{+}]\ep=0
\eea
where $\Theta=(\g_{x^1}\g_{x^2}+\g_{x^3}\g_{x^4})$ and all small $\g$ matrices 
are undressed. Now there are two kind of solutions of the 
above equations. One corresponds to
taking $\g_{+}\p=0$, these are called `standard' Killing spinors.\footnote{ We 
are in the frame where $(\g_{+})^2=(\g_{-})^2=0$ and 
$[\g_{+},\g_{-}]_{-}=2$ 
and the projector is $\g_{-}\g_{+}$. Note that $\g_{+}$ is not a 
projector.} This condition keeps 16 spinors out of the set of total 32.
 For these spinors except   
$\pa_{-}\ep+\cdots=0$ all other equation can be trivially satisfied. 
These sixteen standard killing spinors are \cite{clp,hull}
\be
\ep= e^{-{m\ov2}\g_9\Theta~ x^{-}}\p ~ , ~~~~~~~~ \g_{+}\p=0 ~.    
\ee
All these spinors depend on $x^{-}$ except those which are annihilated by 
$\Theta$.
Rest of the 
 Killing spinors, for which $\g_{+}\chi\ne 0 $, are  usually called as 
`supernumerary'  Killing spinors. 
These can be constructed out of the sixteen spinors $\chi$ 
 with a condition  $\Theta\chi=0$. For these spinors all but the 
equations $\pa_a\ep+\cdots=0$ and $\pa_{-}\ep+\cdots=0$ are to be solved.
These solutions can be written in the simplified form as
\be
\ep= (1+{m\ov4}\g_9\Theta\g_{x^a}\g_{+}x^a)\chi,~~~~~~~~~\theta\chi=\chi~,
\ee
with $\theta=\g_{x^1}\g_{x^2}\g_{x^3}\g_{x^4}$. Due to the condition 
$\theta\chi=\chi$  half of the $\chi$'s are vanishing and  we are left 
with only 8 supernumerary Killing spinors. Thus  total number of standard 
and the supernumerary 
Killing spinors is 24. All these Killing spinors 
are independent of the transverse $y^\a$ and $y^9$ coordinates . Since 
$\p=\p_{+}+\p_{-}$, with $\t~\p_{\pm}=\pm\p_{\pm}$, then 
$\Theta$ 
will automatically annihilate half of the 16 
standard Killing spinors. These 8 standard Killing spinors will then be 
independent of the $x^{-}$. Thus 8 standard Killing spinors as well as  all 
the supernumerary Killing spinors are  
independent of $x^{-}$. In conclusion, supersymmetry of the pp-wave 
background is  more than that of the corresponding 
$AdS_3\times S^3\times T^5$ background. 
Thus there is enhancement of supersymmetries of $AdS$ spacetime in the Penrose 
limit.      
In the next section we shall embed  M5-branes 
in this pp-wave background and we will 
find that the number of supersymmetries is halved. 

\subsection{String coupling: }
Before closing this review let us also describe the importance of 
the constant $c$ in the function $W$  which we have incorporated 
in our backgrounds by simply using modified 
 coordinate change rules in \eqn{3a}. 
Consider the  pp-wave solution \eqn{wave1aa} which can be dimensionally 
reduced along any of the isometry directions. In particular the reduction along 
$x^{-}$ would give following deformed D0-branes with 16 supercharges
\bea\lll{DO}
&& ds^2_{10}= -W^{-{1\ov2}}dx^+dx^+ +W^{{1\ov2}} \bigg[ 
\sum_{a=1}^{4}(dx^a)^2+ \sum_{\a=1}^{4}(dy^\a)^2+(dy^9)^2\bigg]\br
&&H_3^{NS}=2 m (dx^1dx^2dy^9+dx^3dx^4dy^9) ,~~~~W=c-m^2\sum_{a=1}^{4} 
(x^a)^2¸\br && 
e^{2\f}= W^{3/2},~~~A_1=W^{-1}dx^+ \ .
\eea
There are sixteen supersymmetries because 16 Killing spinors which are 
independent of 
the $x^{-}$ coordinate in \eqn{wave1aa} survive after compactification.
There is a constant flux of NS-NS 3-form in these D0-brane solutions that 
leads to the deformation.
Note that $x^+$ coordinate of pp-wave after compactification plays the role of 
the time coordinate. From \eqn{DO} it is clear that for string coupling 
and the geometry to be well defind 
$W$ must be non-negative. The constant $c$ is related to the background 
value of the string coupling $g_s$ at the origin $x^a=0$. 
 Although the pp-wave solution 
\eqn{wave1aa}
holds good without the constant $c$ (as it can be absorbed by the 
shifts $dx^+\to~dx^+-c/2 dx^{-}$), but it becomes an 
important parameter after compactification along $x^-$ 
when we try to make contact with D0-branes in 
\eqn{DO}. A reduction of 
\eqn{wave1aa} along any of the transverse coordinates $y^\a,~y^9$ would 
give rise to pp-wave solutions of type IIA string theory. 

A similar conclusion follows if we consider
maximally supersymmetric Hpp-wave background  in type IIB theory \cite{blau}
which upon T-duality along $x^{-}$ (although this has no Killing 
isometries along $x^-$) describes deformed type IIA fundamental strings in 
presence of constant $F_4$-flux
\bea
&& ds^2_{10}= W^{-{1}}\left(-dx^+dx^+ +dx^{-}dx^{-}\right)+  
\sum_{a=1}^{8}(dx^a)^2.\br
&&B_{+-}=W^{-1} ,~~~~F_4=2m(dx^1dx^2dx^3dx^4+dx^5dx^6dx^7dx^8),
\br && 
e^{2\f}= 
W^{-1},~~~W=c-m^2\sum_{a=1}^{8} x_a^2 \ .
\eea
Here again $W$ gets related to the string coupling in type IIA string theory.

\section{M5-branes in pp-wave background}
Our objective in this paper is to construct solitonic M5-brane solutions in 
M-theory in supersymmetric pp-wave backgrounds. There can be many ways to 
construct such  solutions, we follow here the most obvious and simple procedure 
which involves first writing down the intersecting M2/M5/M5 brane 
configuration \cite{boonstra}
\bea
&& ds^2_{11}= 
f^{-{2\ov3}}H_1^{-{1\ov3}}H_2^{-{1\ov3}}(-dt^2+dz^2)+
f^{-{2\ov3}}H_1^{{2\ov3}}H_2^{{2\ov3}} (dy^9)^2+\br &&~~~~~~~~~~~~
f^{{1\ov3}}H_1^{{2\ov3}}H_2^{-{1\ov3}} \sum_{a=1}^4(dx^a)^2+
f^{{1\ov3}}H_1^{-{1\ov3}}H_2^{{2\ov3}} \sum_{\a=5}^8(dy^\a)^2\ 
,\br
&&G_4=[d~f^{-1}dtdz+\ast dH_1+\ast dH_2]dy^9
\label{2a}\eea 
where Hodge $\ast$ operations are defined over 4-dimensional
 flat  $x^a$ and $y^\a$ coordinate patches 
respectively.\footnote{ A similar construction was done in \cite{kumar} 
for NS5 branes in pp-wave background.} The 
harmonic functions satisfy the equation
\be
(H_2\nabla_x + H_1\nabla_y)f=0,~~~H_1=1+{Q_1\ov 
x^2},~~~H_2=1+{Q_2\ov y^2}   
\lll{har}\ee
where $\nabla$'s are  Laplacians defined over two four-plane. 
For this $f=H_1 H_2$ and 
$f=H_1$ are the two most obvious solutions of  \eqn{har}. For the 
latter case \eqn{2a} becomes
\bea
&& ds^2_{11}= 
H_2^{-{1\ov3}}\left( H_1^{-{1}}(-dt^2+dz^2)
+H_1\sum_{a=1}^4(dx^a)^2\right)+
H_2^{{2\ov3}} \left(\sum_{\a=5}^8(dy^\a)^2+(dy^9)^2\right)\ 
,\br
&&G_{4}=[d H_1^{-1}dtdz+\ast dH_1+\ast dH_2]dy^9
\eea 
Above solution has a near horizon limit $x\to 0$ in which the 
solution becomes M5-brane with anti-de Sitter world-volume
\bea
&& ds^2_{11}= 
H_2^{-{1\ov3}}\bigg[ AdS_3(Q_1)\times S^3 (Q_1)\bigg]+
H_2^{{2\ov3}} \left(\sum_{\a=5}^8(dy^\a)^2 +(dy^9)^2\right)\ ,\br
&&G_{4}=[2 Q_1 \O(AdS_3) -2Q_1 \O(S^3)+\ast dH_2]dy^9
\lll{epa}\eea 
where $ \O(M)$ represents the volume form of unit $M$ space.
Note that $AdS_3(Q_1)$ and $S^3(Q_1)$ have equal size and is given by $Q_1$. 
This solution preserves 
eight supersymmetries. We shall write down the corresponding 
Killing spinors in the next subsection. Thus there is no enhancement 
(doubling) of 
the supercharges in the near horizon limit $x\to 0$. 
This solution  represents a 
solitonic M5-brane which has a world volume wrapped on $AdS_3\times 
S^3$ and is asymtotically ($y\to\infty$) the $AdS_3\times S^3\times T^5$ 
spacetime which has 
16 supersymmetries. 
There is an over all isometry direction $y^9$ and therefore these
M5-branes are different from the usual ones. These are  like 
(smeared) circularly symmetric M5-branes. Also this construction is some what
 unique and we have 
checked that with  $AdS_3\times S^3$ world volume there are no solutions 
which depend upon all the five transverse coordinates.

Having obtained such a configuration of M5-branes, we would like to find out 
what will happen to these solutions in the Penrose limit. The Penrose limit 
of the asymptotic $AdS_3\times S^3\times T^5$ geometry is given in 
\eqn{wave1aa}.
For \eqn{epa} we  take the Penrose (scaling) limit in which scaling parameter $\l\to 0$ 
and is accompanied with the scalings 
$ ~Q_2\to \l^2 Q_2,~y^\a\to \l y^\a,~y^9\to \l y^9$. In order to obtain 
nontrivial scaling limit we have to first express the $AdS_3\times S^3$ 
part of the spacetime in suitable light-cone coordinates $U,~V, X^a$ as 
done in \cite{blau1}, follow it with  the scalings 
$U\to~u,~V\to\l^2~v,~X^a\to\l~ x^a$ and then take the limit $\l\to 0$. 
Using coordinate change rules \eqn{3a}  
we get \bea 
&& ds^2_{11}=\l^2\bigg[ 
H_2^{-{1\ov3}}\left( 2dx^+dx^{-}+W~(dx^{-})^2+\sum_{a=1}^4(dx^a)^2\right)+
H_2^{{2\ov3}} \left(\sum_5^8(dy^\a)^2+(dy^9)^2\right)\bigg]\ ,\br
&&G_4=\l^3[2 Q_1dx^{-}(dx^1dx^2+dx^3dx^4)+\ast 
dH_2]\wedge dy^9,\br
&&H_2=1+{Q_2\over y^2},~~~~~W=1-Q_1^2\sum_{a=1}^4(x^a)^2  \ .
\lll{wave1a}\eea 
Since under this limit 11-dimensional supergravity action scales 
homogeneously, the background 
\eqn{wave1a} represents a solution of supergravity equations even if we 
set $\l=1$ in this solution. We will also set $Q_1=1$ from 
now on for simplicity.     

\subsection{Killing Spinors}

We shall now find out the amount of supersymmetry 
preserved by the M5-brane background in eq.\eqn{wave1a}. It is 
important to know exactly the Killing spinors since they are
 crucial for determining the supersymmetry content of the theory upon 
compactification. Let us first write down the 
Killing spinors for 5-brane in $AdS_3\times S^3\times T^5$ 
background without giving  details. The Killing spinors for the 
background \eqn{epa} are
\bea && \ep=H_2^{-{1\ov12}} F_{AdS} F_{S}~ \ep_0 \br
&&F_{AdS}=[e^{r\over2}P_{+}+(e^{-{r\over2}}+ e^{{r\ov2}}(t \g_t+z 
\g_z)\g_r)P_{-}] \br
&&F_{S}=e^{-{\t\ov2}\g_\f\g_\p\g_{y^9}}e^{-{\f\ov2}\g_\f\g_\t}
e^{-{\p\ov2}\g_\p\g_\f} ,
\eea
where all small $\g$-matrices are undressed, $P_{\pm}= 
{1\ov2}(1\pm\g_t\g_z\g_{y^9})$, with $\g_t,~\g_z,~\g_r,~
\g_\t,~\g_\f,~\g_\p$ 
being along   
$AdS_3\times S^3$ in the same order (we have set $Q_1=1$).\footnote{ 
Anti-de Sitter metric can be written as 
$e^{2r} (-dt^2+dz^2)+dr^2$ 
and $S^3$ line element is taken to be $d\t^2+sin^2\t d\f^2+ sin^2\t sin^2\f 
d\p^2.$} 
The constant spinor $\ep_0$ satisfies the constraints 
\be
 \g_{y^5}\g_{y^6}\g_{y^7}\g_{y^8}\g_{y^9}\ep_0=-\ep_0, ~~~
 \g_t\g_z\g_r\g_\t\g_\f\g_\p\ep_0=-\ep_0.
\label{ops}\ee
 These twin conditions 
break the supersymmetries to one-quarter. The two sets of operaters in 
\eqn{ops} commute with 
each other. These operators also commute with $F_{AdS}$ and $F_{S}$ as 
well. When the charge of M5-branes vanishes ($i.e.~ Q_2=0$) the first 
condition drops 
out and the supersymmetry is increased to sixteen. Thus embedding of the branes 
in $AdS_3\times S^3\times T^5$ explicitly breaks  half of the 
supersymmetries. 
This should also be the case when the five-branes are 
considered in pp-wave background. In particular it is interesting to know 
what  happens to the standard and the supernumerary Killing spinors when 
five-branes are embedded in pp-wave background.

We again write down the tangent space metric as $ds^2=2e^{+}e^{-}+ 
e^a~e^a+e^\a~e^\a+e^9e^9$,
where tangent space indices are taken same as the space-time indices. 
The basis elements are now given by 
$e^+=H_2^{-{1\ov6}}(dx^{+}+{W\ov2} 
dx^{-}),~e^{-}=H_2^{-{1\ov6}}dx^{-},~e^a=H_2^{-{1\ov6}}dx^a,~e^\a=
H_2^{{1\ov6}}dy^\a,~ e^9=H_2^{{1\ov3}}dy^9$. Correspondingly 
 the spin connections are 
$$\o^{+a}={1\ov2}\partial_a W 
dx^{-},~\o^{+\a}=-{1\ov6}\pa_\a H_2 H_2^{-{3\ov2}}(dx^{+}+{W\ov2}dx^{-}),~
\o^{-\a}=-{1\ov6}\pa_\a H_2 H_2^{-{3\ov2}}dx^{-},$$ $$\o^{a\a}=-{1\ov6}\pa_\a 
H_2 
H_2^{-{3\ov2}} dx^a,~\o^{\a\b}={1\ov3}\pa_\a H_2 
H_2^{-1}dy^\a,~\o^{9\a}={1\ov3} \pa_\a H_2 H_2^{-1}dy^9.$$
With these spin connections we solve 
for the Killing equations in \eqn{killing}. We find 
for \eqn{wave1a} the standard Killing spinors ($\g_{+}\p=0$) are 
given by
\be
\ep= H_2^{-{1\ov12}}(y) e^{-{1\ov2}\g_9\Theta~ x^{-}}\p ~ , ~~~~ 
~~\bar\G \p=-\p ,    
\lll{3a1}\ee
which are 8 in number while the supernumerary ones ($\g_{+}\p\ne0$) are
given by
\be
\ep= H_2^{-{1\ov12}}(y) 
(1+{1\ov4}\g_9\Theta\g_a\g_{+}x^a)\chi,~~~~~~~~~\Theta\chi=0~,~~~
\bar\G\chi=-\chi ,
\lll{3b}\ee
with $\bar\G=\g_{y^5}\g_{y^6}\g_{y^7}\g_{y^8}\g_{y^9}$. 
Note that the  $\Theta$ and $\bar\G$ commute with each other which is crucial.
Thus the number of supernumerary killing spinors is only four and the 
total number of the Killing spinors for  M5-brane embedded in pp-wave 
background  becomes twelve. All of these  spinors 
are independent of the coordinate $y^9$ only.  All supernumerary Killing 
spinors and half of the standard ones are also independent of the $x^{-}$ 
coordinate. That is total of 8 Killing spinors are independent of $x^{-}$.
These will survive if we  compactify the $x^{-}$ direction on a circle. 

In conclusion we have shown that both standard as well as the 
supernumerary  
Killing spinors exist for the five-brane background \eqn{wave1a} but 
their numbers are reduced by 
half due to the additional condition $ 
\g_{y^5}\g_{y^6}\g_{y^7}\g_{y^8}\g_{y^9}\ep=-\ep$
in the transverse space. This condition was absent for pp-wave 
background in \eqn{wave1aa}.

The compactification of \eqn{wave1a} along the 
$y^9$ coordinate will give NS5-branes in pp-wave background of type IIA, which 
has been 
considered by Kumar et.al.  \cite{kumar}. All the Killing spinors in 
eqs. \eqn{3a1} and \eqn{3b} will 
survive in this compactification. Thus we have 
 provided the 
M-theory relationship for  NS5-branes in pp-wave backgrounds having 24 
supersymmetries. The existence of half supersymmetric (with 16 susy) 
D3, D5 and D7-branes 
was recently shown by Dabholkar et.al. \cite{dabhol} where the branes are 
embedded in maximally supersymmetric Hpp-wave backgrounds \cite{blau}. However, 
our M5-branes  only preserve 12 supersymmetries which is  half of the 
amount preserved by the asymptotic pp-wave background. It would be 
interesting to see if half supersymmetric M5-branes can be embedded into
maximally supersymmetric Mpp-wave backgrounds.

\leftline{\bf Acknowledgments}
{I would like to thank A. Micu and S. Theisen for useful discussions. 
This work is supported by AvH (the 
Alexander von Humboldt foundation).}

\end{document}